\documentclass[letterpaper]{aa}  
\usepackage{natbib}
\bibpunct{(}{)}{;}{a}{}{,}
\usepackage{graphicx}
\usepackage{txfonts}
\begin{document}
 \title{Photon Dominated Region Modeling of Barnard~68}

   \author{J. L. Pineda\thanks{Member of the International Max
          Planck Research School (IMPRS) for Radio and Infrared Astronomy at
          the University of Bonn and Cologne}
          \and
          F. Bensch
          }

  \offprints{J. L. Pineda \email{jopineda@astro.uni-bonn.de}}

  \institute{Argelander Institut f\"ur Astronomie, Universit\"at Bonn, Auf dem
    H\"ugel 71, D-53121 Bonn, Germany}

   \date{}

   \abstract {} { We use the Barnard~68 dark globule as a test case for a
     spherically symmetric PDR model exposed to low-UV radiation fields.  With
     a roughly spherical morphology and an accurately determined density
     profile, Barnard~68 is ideal for this purpose.  The processes governing
     the energy balance in the cloud surface are studied in detail.}  { We
     compare the spherically symmetric PDR model by St\"orzer, Stutzki \&
     Sternberg (1996) to observations of the three lowest rotational
     transitions of $^{12}$CO, $^{13}$CO $J = 2 \to1$ and $J = 3 \to 2$ as
     well as the [C\,{\sc i}] $^3$P$_1$$\to$$^3$P$_0$ fine structure
     transition. We study the role of Polycyclic Aromatic Hydrocarbons (PAHs)
     in the chemical network of the PDR model and consider the impact of
     depletion as well as of a variation of the external FUV field.  }  { We
     find it difficult to simultaneously model the observed $^{12}$CO and
     $^{13}$CO emission.  The $^{12}$CO and [C\,{\sc i}] emission can be
     explained by a PDR model with a external FUV field of 1-0.75\,$\chi_0$,
     but this model fails to reproduce the observed $^{13}$CO by a factor of
     $\sim$2.  Adding PAHs to the chemical network increases the [C\,{\sc i}]
     emission by 50\% in our model but makes [C\,{\sc ii}] very faint. CO
     depletion only slightly reduces the $^{12}$CO and $^{13}$CO line
     intensity (by $\lesssim$10\% and $\lesssim$20\%, respectively).
     Predictions for the [C\,{\sc i}] ${^2}$P$_{3/2}$$\to$${^2}$P$_{1/2}$,
     [C\,{\sc i}] $^3$P$_2$$\to$$^3$P$_1$ and $^{12}$CO $J= 5\to4$ and 4$\to$3
     transitions are presented.  This allows a test of our model with future
     observations (APEX, NANTEN2, HERSCHEL, SOFIA).} {}

  \maketitle
%

\section{Introduction}
\label{sec:introduction}

The understanding of Photon Dominated Regions (PDRs) is of great interest as
they account for much of the millimitre, sub-millimitre and far-infrared line
and continuum radiation from the molecular interstellar medium (ISM) of nearby
star forming regions to clouds in the diffuse Galactic radiation field. Many
numerical codes have been developed in order to model their emission and to
understand the physical and chemical processes governing them
\citep{HollenbachTielens99}. Typically, PDR models use a physical structure
(morphology, density profile) to calculate a self-consistent solution for the
chemistry (abundance of species) and the energy balance (heating and cooling).
Practically, the numerical codes differ in the scope of the chemical networks
and their reaction rates as well as the degree of sophistication with which
the heating, cooling and the radiative transfer are calculated.  In many cases
the goal of the PDR modelling is to constrain the physical properties of the
line-emitting region (for example, its density and column density) and the
strength of the external far-ultraviolet (FUV) field.  PDR models
  successfully account for much of the millimitre, sub-millimitre and
  far-infrared line and continuum emission from Galactic and extra-galactic
  PDRs \citep{Hollenbach97,HollenbachTielens99}.

The coupling of the energy balance (heating \& cooling) and the chemistry
makes PDR modelling not a straightforward task, however. In particular, a
large set of non-linear equations needs to be solved in order to obtain
abundance profiles of species. Small changes in the initial elemental
abundances or the reaction rates of individual reactions within their
sometimes large error margins can result in large variations of the abundance
profiles of individual species \citep{Roellig06a}.  In some regions of the
parameter space, more than one steady-state solution to the chemical network
may exist \citep[bistability, e.g.][]{LeBourlot93}.  Additional complications
arise from intrinsic variations of published PDR models which are noted even
when identical source parameters and chemical reaction rates are used. The
benchmarking study \citep{Roellig06a} has shown that even for constant density,
plane-parallel models, the emerging line intensity of different numerical
codes vary by up to an order of magnitude for some of the fine structure
transitions.  These discrepancies likely reflect the different numerical
methods which are used to solve the chemical structure as well as the
numerical implementation of physical processes.   The calibration of PDR
  models cannot be easily done since the observed clouds generally have more
  complex morphologies than assumed in most PDR models.  It is therefore
desirable to study clouds with a well-known geometry (density profile) which
can be used to calibrate PDR models.  For example, numerous PDR models
  have been made for the Orion Bar because of its approximate plane-parallel
  geometry \citep[e.g.][]{TielensHollenbach85,HollenbachTielens99}. 

Barnard~68 is an example where the density profile is constrained to a very
high precision \citep{Alves01}, and its geometry is well matched to a
spherically-symmetric PDR model. With no young stars nearby, the external FUV
field is low, of the order of the diffuse Galactic radiation field.  Due to
its proximity to the Ophiuchus complex, we adopt a distance to the cloud of
125~pc \citep{deGeus89}. Using near-infrared extinction techniques,
\citet{Alves01} derived an extinction map that was used to constrain the radial
density profile of the cloud, suggesting that Barnard~68 is consistent with a
pressure-confined self-gravitating cloud near equilibrium, a Bonnor-Ebert
sphere. Avoiding the uncertainly concerning the geometry and density structure
we are able to study in detail the impact of the chemistry in the predicted
line emission of the model.

Here, we present observations of the three lowest rotational transitions of
$^{12}$CO, the [C\,{\sc i}] $^3$P$_1 \to ^3$P$_0$ fine structure transition of
neutral carbon, $^{13}$CO $J = 2\to1$, and $^{13}$CO $J = 3\to2$ and compare
them to the KOSMA-$\tau$ spherical PDR model originally published by
\citet{Stoerzer96}. The observations are summarized in Section
\ref{sec:observations}. Section \ref{sec:pdr-model} describes the employed
model, and the results are discussed in Section \ref{sec:results}.  In
Section~\ref{sec:Discussion} the gas heating and cooling is studied and
predictions of the intensity for other important cooling lines are made. A
summary is given in Section~\ref{sec:conclusions}.


   \begin{figure}[t]
     \label{colden}
     \centering
   \includegraphics[width=7cm,angle=-90]{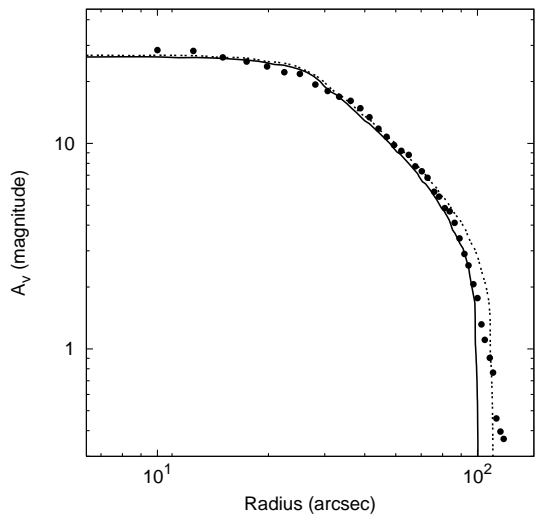}
      \caption{ Extinction profile for Barnard\,68.
        Filled dots represent the measured A$_\mathrm{v}$ calculated by
        \citet{Alves01}, excluding the south-east prominence seen in the
        visual extinction map.  The solid line gives the line-of-sight
        extinction profile versus radius for the power-law density profile
        described in Section~\ref{sec:reference-model}.  The dashed line is
        the same column density profile for a model with the same density
        power-law index but a 10\% larger radius.  }\label{colden}
   \end{figure}

%

\section{Observations}
\label{sec:observations}

%
\begin{table*}[t]
\caption{Spectral Line Observations Toward Barnard~68}             
\label{tab:observations}
\centering                          
\begin{tabular}{l c c c  c c c c }  
\hline\hline                 
Telescope & Transition & Frequency &\# of Positions  & Sampling & $\theta_{\textrm{mb}}$  &
$\eta_\textrm{mb}$ & $\Delta$v \\   
& & GHz  & [arcsec]  & [arcsec] &  & [km s$^{-1}$]     \\    

\hline                        
SWAS & [C\,{\sc i}] $^3$P$_1\to ^3$P$_0$ & 492.160 & 1 & $-$   &  258 & 0.9 & 0.623  \\
Mopra & $^{12}$CO $J = 1 \to 0$ & 115.271 & 144  & 15   &   33 & 0.42 & 0.08 \\ 
KOSMA & $^{12}$CO $J = 2 \to 1$ & 230.538 &49 &  60 & 130& 0.76 & 0.1\\  
KOSMA & $^{12}$CO $J = 3 \to 2$ & 345.795 & 49 &  60& 80 & 0.78 & 0.294\\
KOSMA & $^{12}$CO $J = 3 \to 2$ & 345.795 &1 &  $-$ & 80 & 0.78 & 0.024\\
CSO$^{1}$ & $^{13}$CO $J = 2 \to 1$ & 220.398 &   65  & 24   & 33 & 0.70 & 0.06\\
CSO$^{1}$ & $^{13}$CO $J = 3 \to 2$ & 330.587 &  144 & 24  &  22 & 0.75 & 0.04\\
\hline   
\multicolumn{7}{l}{$^1$ Published by \citet{Bergin06}.}                                
\end{tabular}
\end{table*}

\begin{table}[t]
\caption{Standard Model Parameters}             
\label{tab:reference}
\centering                          
\begin{tabular}{l c}        
\hline\hline                 
Cloud surface density (cm$^{-3}$) & 2.01 $\times$ 10$^{4}$ \\
Cloud radius (cm)& 1.875 $\times$ 10$^{17}$   \\
Exponent of the density power-law &  1.96 \\
Line Doppler width (km~$s^{-1}$) & 0.11 \\
\hline                                   
\end{tabular}
\end{table}

\begin{table}[t]
\caption{Fractional Abundances$^\mathrm{a}$}             
\label{tab:fractional}
\centering                          
\begin{tabular}{l c}        
\hline\hline                 

He  & 0.1  \\
C  & 1.32 $\times$ 10$^{-4}$  \\
O  & 2.91 $\times$ 10$^{-4}$     \\
N  & 8.5 $\times$ 10$^{-5}$ \\
$^{13}$C  & 2.08  $\times$ 10$^{-6}$  \\
S  & 1.87 $\times$ 10$^{-6}$      \\
Mg  & 5.12  $\times$ 10$^{-6}$     \\
Fe  & 6.19 $\times$ 10$^{-6}$      \\
Si  & 8.21 $\times$ 10$^{-7}$     \\
PAH$^\mathrm{b}$ & 1 $\times$ 10$^{-7}$      \\
\hline                                   
$^\mathrm{a}$ Abundance relative to H nuclei. \\
$^\mathrm{b}$ Not included in the reference model.
\end{tabular}
\end{table}

Observations were made for the three lowest rotational transitions of CO and
the [C\,{\sc i}] $^3$P$_1\to^3$P$_0$ fine structure transition of neutral
carbon in Barnard~68.  Additionally, we use the $^{13}$CO $J =3 \to 2$ and $J
=2 \to 1$ rotational transitions observed with the CSO telescope and published
by \citet{Bergin06}.  Using these transitions we deliberately limit ourselves
to surface tracers, since the goal of this paper is to model of the PDR
emission of a cloud in the diffuse Galactic radiation field.

The high signal-to-noise [C\,{\sc i}] $^3$P$_1$$\to$$^3$P$_0$ observation was
made in 2001 with the Submillimiter Wave Astronomy Satellite
(SWAS\footnote{See \cite{Melnick00} for additional details about SWAS.}) and
has an angular resolution of 4\farcm3.  It is pointed at a position ($\Delta
\alpha=-$0\farcm094, $\Delta \delta=$+0\farcm2) offset from the column density
peak of Barnard~68 at $\alpha$ = 17$^{\tiny \textrm{h}}$22$^{\tiny
  \textrm{m}}$38\fs6 and $\delta$ = $-$23\degr49\arcmin46\farcs0 (J2000)
\citep{DiFrancesco02} and covers essentially the whole cloud.  The r.m.s noise
$\Delta T_\mathrm{mb}$ is 23~mK per velocity channel ($\Delta$v$=0.623$
km~s$^{-1}$).  A 3rd-order polynomial is fitted and subtracted from the
spectra.  SWAS simultaneously observes the transitions of four species,
[C\,{\sc i}] $^3$P$_1 \to ^3$P$_0$ at 492~GHz, $^{13}$CO $J = 5\to4$ at
550.9~GHz, H$_2$O at 556.0~GHz, and O$_2$ at 487.2~GHz. No emission is
detected for the latter three species, however, and \citet{Bergin02} give the
resulting upper limits on the H$_2$O and O$_2$ column densities.  The r.m.s
noise of the $^{13}$CO $J = 5\to4$ spectrum is 13~mK per velocity channel.

The $^{12}$CO $J = 1\to0$ (115~GHz) rotational transition was observed in June
2005 with the Mopra 22-m telescope located in Australia.  A
6\arcmin\,$\times$ 6\arcmin~area was mapped using the on-the-fly mode.
The data were smoothed to an angular resolution of 1\arcmin~on a grid with a
30\arcsec~spacing.  The resulting spectra typically have a r.m.s noise $\Delta
T_{\tiny \textrm{mb}}$ better than 0.5~K per velocity channel
($\Delta$v$=0.08$ km~s$^{-1}$). A straight line was fitted and subtracted from
each spectrum.

   \begin{figure*}[t]
     \label{ci_swas}
     \centering
   \includegraphics[width=\textwidth]{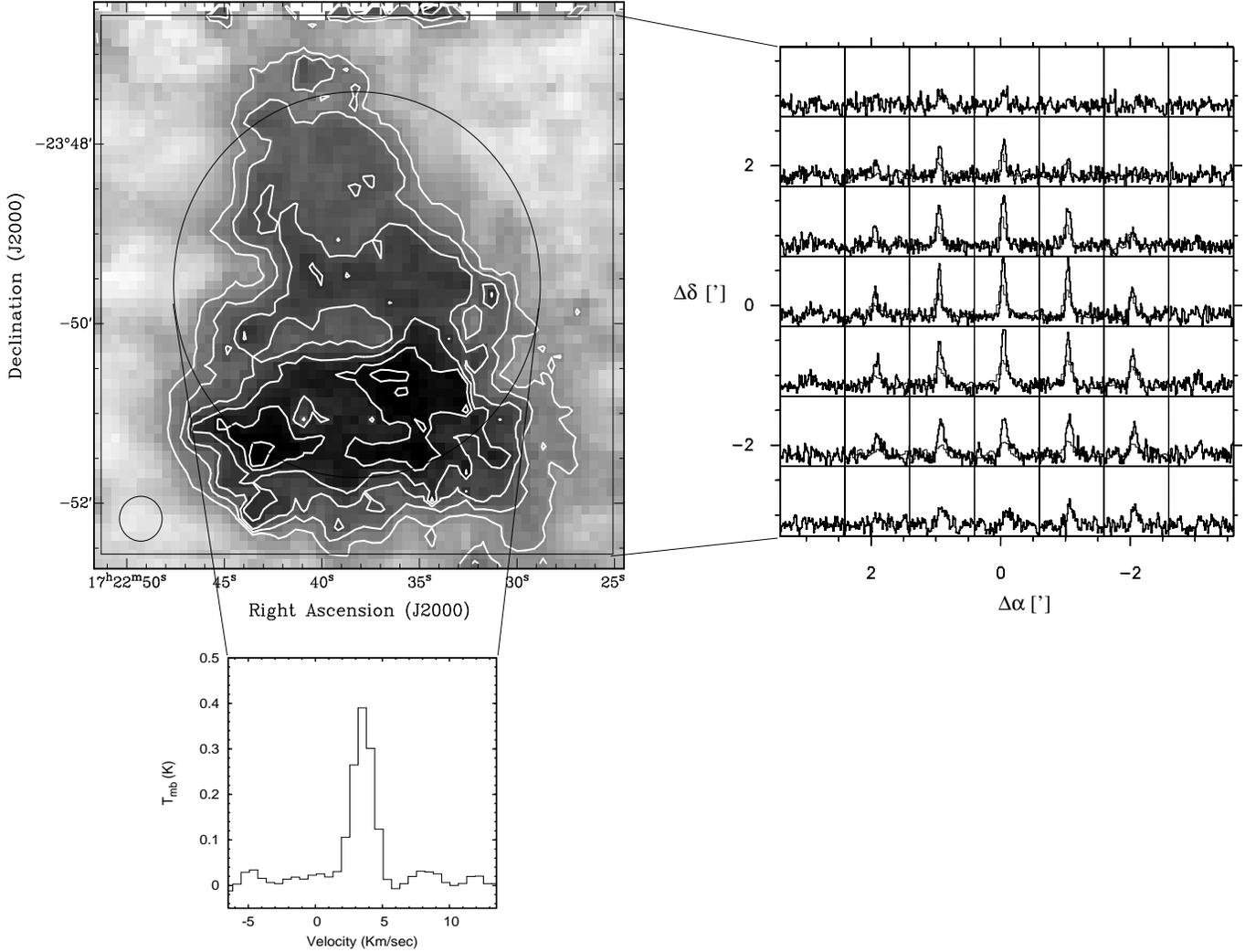}
      \caption{ {\it Upper left:} Map of the $^{12}$CO $J =1 \to 0$ emission of Barnard~68. The map is
        shown in its original resolution of 33\arcsec. The contours run from
        40\% to 90\% of the peak intensity (72.6 K km~s$^{-1}$) in steps of
        10\%.  {\it Upper right: } Spectra of the $^{12}$CO $ J= 2 \to 1$
        (thick lines) and $ J = 3 \to 2$ (thin lines) transitions with a
        resolution of 130\arcsec. The velocity range is v $ = 0 \ldots
        8$~km~s$^{-1}$, and the main-beam temperature shown is $T_\mathrm{mb}
        = -1 \ldots 5$~K .
         The offset is given in arcminutes and are relative to the peak of the
        dust emission at $\alpha$ = 17$^{\tiny \textrm{h}}$22$^{\tiny
          \textrm{m}}$38\fs6 and $\delta$ = $-$23\degr49\arcmin46\farcs0
        (J2000). {\it Lower left: }
        [C\,{\sc i}] $^3$P$_1 \to ^3$P$_0$ emission toward the central
        position of Barnard 68. The 4\farcm3 beam size of SWAS covers a large
        fraction of the cloud.}\label{ci_swas}
   \end{figure*}

   The $^{12}$CO $J = 2\to1$ (230~GHz) and $^{12}$CO $J = 3\to2$ (345~GHz)
   transitions were mapped simultaneously with the 230/345~GHz dual-channel
   receiver at the KOSMA 3-m telescope toward 49 positions, spaced by
   1$\arcmin$.  For the $^{12}$CO $J = 2\to1$ transition 90\% of the spectra
   have a r.m.s noise $\Delta T_{\tiny \textrm{mb}}$ smaller than 0.3~K per
   velocity channel ($\Delta$v$=0.1$ km~s$^{-1}$). The $^{12}$CO $J = 3\to2$
   spectra are smoothed to the 130\arcsec\,angular resolution of the 230~GHz
   beam, and 90\% of the resulting spectra have a r.m.s noise $\Delta T_{\tiny
     \textrm{mb}}$ smaller than 0.5~K per velocity channel ($\Delta$v$=0.294$
   km~s$^{-1}$).  Typically, a sinusoidal or 3rd-order polynomial was fitted
   and subtracted from each spectrum.  Because of a small misalignment of both
   receiver channels the pointing of the 345~GHz beam is offset by
   $\Delta$AZ=$-$7\arcsec~ and $\Delta$EL=$+$28\arcsec~ in horizontal
   coordinates relative to the lower frequency beam.  The observations were
   completed within a relatively short period of time during the transit of
   the source.  Thus, the offset in the horizontal system translates into an
   offset of approximately $\Delta \alpha$ = $-$7\arcsec\,and $\Delta
   \delta$=$-$28\arcsec\,in equatorial coordinates, which was corrected in the
   final map.  We estimate the accuracy of this correction to be within
   $\sim$15\arcsec.  The central position was also observed in the
     $^{12}$CO $J = 3\to2$ line using the high-resolution spectrometer (HRS).
     The calibration of the KOSMA data was regularly checked and corrected
     using a source of known intensity (DR21).  The observed spectra are
   shown in Figure \ref{ci_swas} and a summary of the observations is given in
   Table~\ref{tab:observations}.

\section{PDR Model}
\label{sec:pdr-model}

We employ the spherically symmetric PDR model originally developed by
\cite{Stoerzer96} (KOSMA-$\tau$ model).  This model uses a spherical cloud
with a truncated power-law density profile and an isotropic FUV radiation
field.  We adopt a density profile of,
$n(r)$=$n_\mathrm{s}(r/r_\mathrm{c})^{-\alpha}$ for 0.3$r_\mathrm{c}$ $\leq$ $
r \leq r_\mathrm{c}$, and constant density, n($r$)=$n_s$(0.3)$^{-\alpha}$ in
the cloud center (r $<$ 0.3$r_\mathrm{c}$), for the present paper.  Here,
$r_\mathrm{c}$ is the cloud radius and $n_\mathrm{s}$ is the density at the
cloud surface.   In our models we adopt a power-law exponent of $\alpha$ =
  1.96 for the density profile, a cloud surface density of
  2$\times$10$^{4}$~cm$^{-3}$, and a cloud radius of 1.9$\times$10$^{17}$~cm.
  This gives a reasonable fit to the measured column density profile for $r
  \lesssim$ 100\arcsec, but does not account for the low-column density gas
  close to the surface at 100\arcsec~$\lesssim r \lesssim$
  120\arcsec~(Figure~\ref{colden}).  We tested the response of the emerging CO
  emission to small variations in the model cloud size. An extreme case is
  shown in Figure~\ref{colden} by the dotted line
  ($r_\mathrm{c}=2.09\times10^{17}~$cm), which matches the observed
  A$_\mathrm{v}$ at 100\arcsec~$\lesssim r \lesssim$ 120\arcsec, but
  overestimates the column density for 70\arcsec~$\lesssim r \lesssim$
  100\arcsec. The latter model produces CO emission which is typically larger
  by only a few percent, however.  The Doppler line-width in the model is
  $b$=0.11~km~s$^{-1}$, derived from the observed $^{13}$CO $J= 2 \to 1$ line
  profile.  The gas temperature measured using NH$_3$ observations toward
  Barnard~68 is $10-16\,$K \citep{Hotzel02b,Bourke95}. This suggests a pure
  thermal line width of $0.08-0.1$~km~s$^{-1}$ for CO.    The chemical
network of the model includes H, He, C, O, N, plus a number of heavier
elements, S, Si, Fe, and Mg.  $^{13}$C is considered in the chemical network,
including the isotope selective reactions for $^{13}$CO.  Self-shielding
  of $^{12}$CO and $^{13}$CO and shielding by H$_2$ against photo-destruction
  is implemented using the shielding factors by \citet{vanDishBlack88}.  A
summmary of the model parameters is shown in Table~\ref{tab:reference} and the
initial fractional abundances are listed in Table~\ref{tab:fractional}.  Line
intensities and line profiles are determined for the [C\,{\sc ii}] and the
[C\,{\sc i}] fine structure transitions and the $^{12}$CO and $^{13}$CO
rotational transitions using the radiative transfer code by \cite{Gierens92}.
For a comparison to the observed data we smoothed the model intensity
distribution and line profiles to the angular and velocity resolution of our
observations.   A summary of the models presented in this publication is
  shown in Table~\ref{tab:summary_models}.

\begin{figure*}[t]
  \centering \includegraphics[width=1.2\textwidth, angle=0]{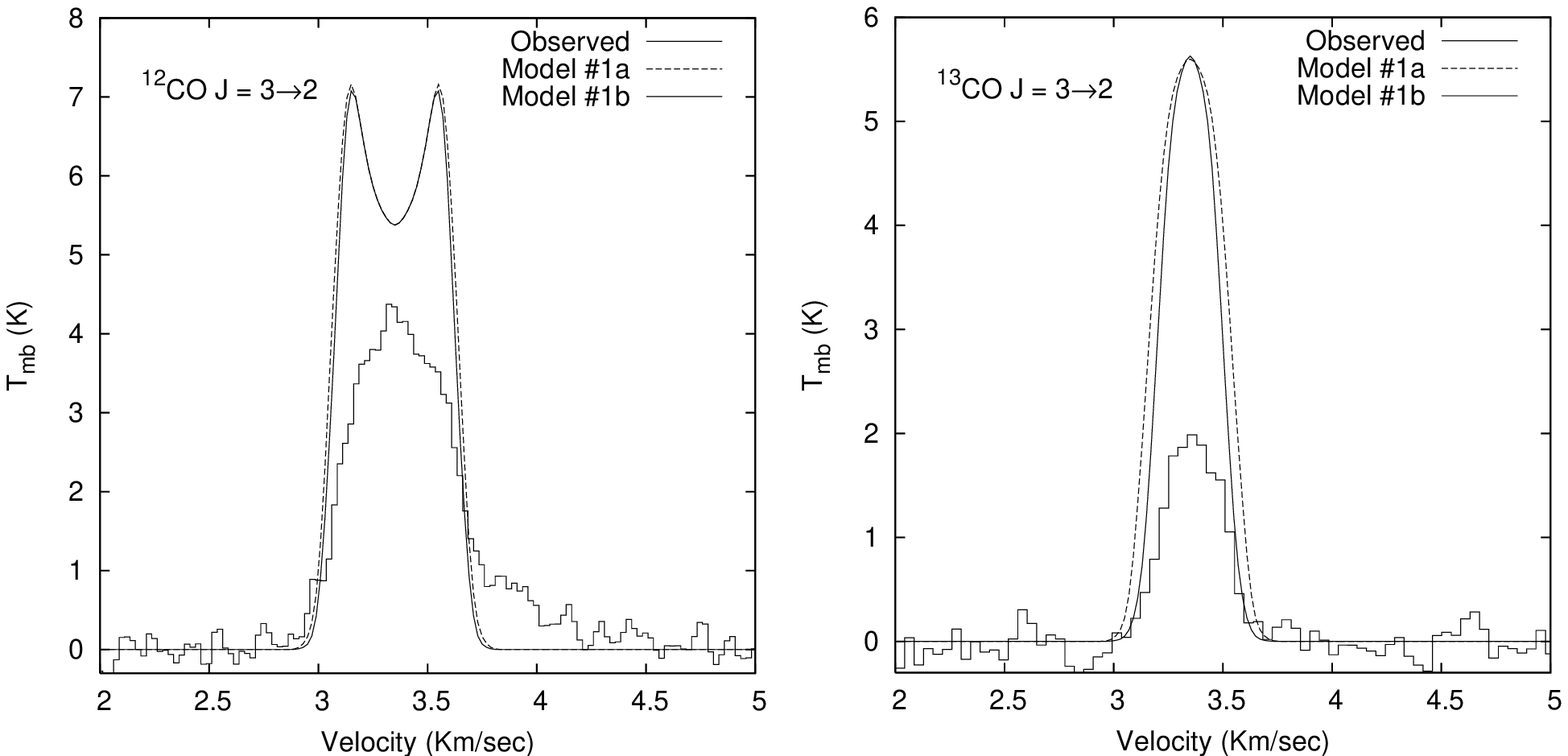}
      \caption{Observed  $^{12}$CO  and $^{13}$CO  $J = 3 \to 2$ line
        profiles. The reference model (Model \#1a) and a model which considers
        the effects of depletion (Model \#1b) are also shown. }
\label{fig:depletion}
\end{figure*}

\section{Results}
\label{sec:results}

%

\begin{table}[t]
\caption{Summary of PDR models}             
\label{tab:summary_models}
\centering                          
\begin{tabular}{c c c c}        
\\[-0.4cm]
\hline\hline
Model  & FUV field  & PAHs & Depletion \\ 
\# & $\chi_0$ &&\\
\hline
 1a & 1.0 & N & N \\
 1b & 1.0 & N & Y \\
 2b & 1.0 & Y & Y \\
 3a & 0.75 & Y & Y\\
 3b & 0.12 & Y & Y \\
\hline                                   
\end{tabular}
\end{table}

\subsection{Reference Model}
\label{sec:reference-model}

 In the following, we define a reference model (Model \#1a) with a
  strength of the external FUV field of $\chi$ = 1.0$\chi_0$, where $\chi_0$
  is the intensity of the mean interstellar radiation field
  (\citealt{Draine78}).  Our modelling goal is to match the observed
  line-integrated intensity and line profiles towards the cloud center as well
  as the azimuthally averaged intensity distribution.
  Table~\ref{table:intensity_comparison} shows the [C\,{\sc i}], $^{12}$CO,
  and $^{13}$CO integrated intensity towards the cloud center (ratio of the
  model simulation to the observations).

Model \#1a reproduces the line-integrated intensity toward the cloud center
for the [C\,{\sc i}] $^3$P$_1 \to ^3$P$_0$, $^{12}$CO $J = 1\to 0$, and $J = 2
\to 1$ transition within 20\%. However, the $^{12}$CO $J = 2 \to 1$ and $J = 3
\to 2$ model line profiles show self-absorption which is not observed
(Figure~\ref{fig:depletion}).

We note that the $^{12}$CO $J = 3 \to 2$ mapping observations for $r<
100$\arcsec~(Figure~\ref{figure:cloudprofile}) have a line intensity larger
than the observation in Figure~\ref{fig:depletion} made toward the center
position.  The $^{12}$CO observations made with a high velocity resolution of
$\sim 0.1$ km s$^{-1}$ show a red line wing, with emission at velocities
between 3.7 and 4.3 km s$^{-1}$ (e.g. left panel of
Figure~\ref{fig:depletion}). This component cannot be modeled in the framework
of a single, spherical clump; for a comparison to the PDR model simulations we
therefore calculated the observed line-integrated intensity excluding this red
line wing. This was not possible for the $^{12}$CO $J = 3 \to 2$ mapping data
because of the insufficient velocity resolution of the variable resolution
spectrometer (VRS), and we expect that the line-integrated intensity of
$^{12}$CO $J = 3 \to 2$ in Figure~\ref{figure:cloudprofile} is systematically
overestimated. We estimate the magnitude of this effect to be of the order of
20\%, based on our velocity-resolved $^{12}$CO observations.  Even with
  the line wing subtracted, however, the integrated intensity of the $^{12}$CO
  $J = 3 \to 2$ HRS spectrum is about 40\% smaller than that for the
  corresponding VRS observations made at a different epoch. The integrated
  intensities of both observations only marginally agree within 3$\sigma$,
  given the error bars of $\sim$16\%.  Despite of a careful check of the
  pointing and calibration of the observations we have to attribute a possible
  systematic error of up to $\sim$40\% to the $J = 3 \to 2$ data.  This is
  likely to be due to the low elevation of the source during the observations
  (Barnard\,68 transits at $\lesssim 20$\degr elevation at the location of the
  KOSMA observatory).  While the $^{12}$CO $J = 3 \to 2$ VRS observations are
  consistent with the model, the HRS spectrum appears to be too weak by a
  factor of 0.5.

The largest discrepancy between the model and the observations is noted for
the $^{13}$CO emission, however, with line-integrated intensities in the model
being larger by 1.9 and 3.2 for the $J = 2\to 1$ and $J = 3 \to 2$
transitions, respectively.  Since the density profile of Barnard~68 is well
constrained, the discrepancies with the observations must have their origin in
the chemical network or the strength of the FUV radiation field.

   \begin{figure*}[t]
     \centering
     \includegraphics[width=1.1\textwidth]{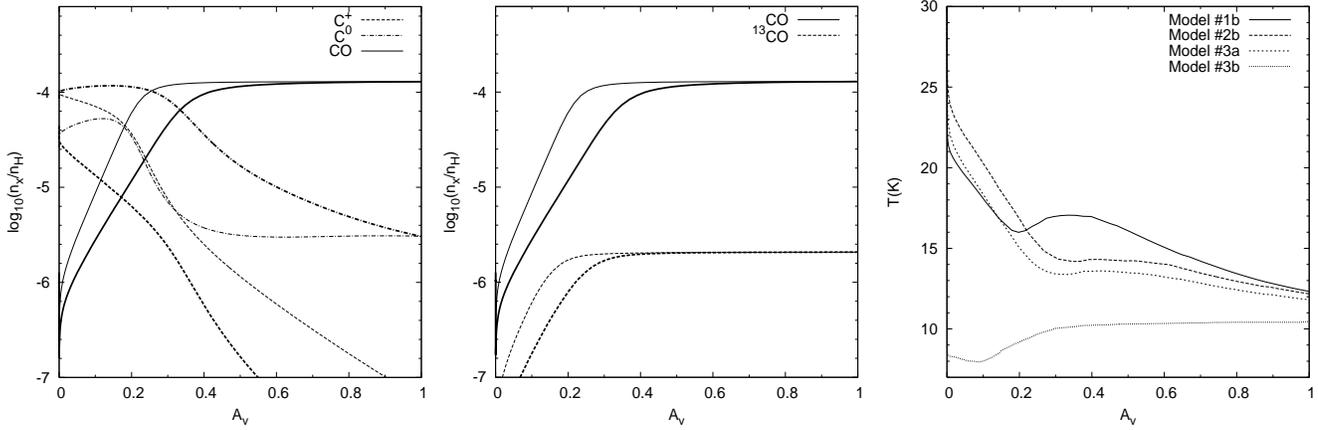}
      \caption{ {\it Left panel} The abundance profile of C$^+$, C$^0$, and CO
        for Model \#1b (thin lines) and Model \#2b (thick lines) are compared.
         {\it Central panel} The abundance profile of CO and $^{13}$CO for
          Model \#1b (thin lines) and Model \#2b (thick lines).  {\it Right
          panel} Gas temperature at the cloud surface as a function of
        A$_\mathrm{v}$ for models \#1b, \#2b,\#3a, and \#3b.}
        \label{figure:2}
   \end{figure*}


\begin{table*}[t]
\caption{Model comparison with observations  toward the Barnard~68 center}  
\label{table:intensity_comparison} 
\centering          
\begin{tabular}{l l c c c c c c c  }
\hline\hline           
Species Transition & Telescope &Resolution (\arcsec)  &$I_\mathrm{obs}^{a}$ &  \multicolumn{5}{c}{$I_\mathrm{model}/I_\mathrm{obs}$} \\
 & & &   & Model \#1a$^{b}$ & Model \#1b &   Model \#2b  &  Model \#3a  &  Model \#3b \\    
\hline    
[C\,{\sc i}] $^3\mathrm{P}_1$$\to$$^3\mathrm{P}_0$ &SWAS & 258&
0.93$\pm$5\% & 0.97 & 0.98 & 1.55 & 1.39 & 0.41 \\
$^{12}$CO $J = 1 \to 0$ & Mopra & 60 & 
0.11$\pm$9\% & 1.08 & 0.99 & 0.87 & 0.82 & 0.51 \\
$^{12}$CO $J = 2 \to 1$ & KOSMA & 130 &
0.47$\pm$2\% & 1.24 & 1.18 & 1.10 & 1.00 & 0.54 \\
$^{12}$CO $J = 3 \to 2$ & KOSMA/HRS$^{b}$ & 80 &
0.72$\pm$ 15\% & 2.26& 2.14 & 2.03 & 1.56 & 0.68\\
$^{12}$CO $J = 3 \to 2$ & KOSMA/VRS$^{b}$ & 80 &
1.41 $\pm$ 16\% & 1.15 & 1.09 & 1.03 & 0.80 & 0.35 \\
$^{13}$CO $J = 2 \to 1$ & CSO & 33 & 
0.19$\pm$5\% & 1.92 & 1.56 & 1.38 & 1.30 & 0.94 \\
$^{13}$CO $J = 3 \to 2$ & CSO & 22 &
0.25$\pm$8\% & 3.19 & 2.57 & 2.13 & 1.95 & 1.27 \\
\hline                               
\multicolumn{9}{l}{$^{a}$ The intensities are in units of 10$^{-7}$
  erg~s$^{-1}$~cm$^{-2}$~sr$^{-1}$. }\\
\multicolumn{9}{l}{$^{b}$ Observations made at different epochs with the
  high-resolution spectrometer (HRS) and the  variable-resolution spectrometer
  (VRS).}\\
\end{tabular}
\end{table*}

\subsection{Depletion}
\label{sec:depletion}

The observed line emission is significantly overestimated by the model for
those transitions which probe somewhat deeper layers into the cloud surface,
most notably the $^{13}$CO transitions.  Observations of C$^{18}$O and
C$^{17}$O in Barnard 68 reveal that CO, among other molecules, is depleted
from the gas phase at A$_\mathrm{v} \gtrsim 5$ due to freezing on dust grains
\citep{Hotzel02,DiFrancesco02,Bergin06}.  In the present model, we study the
impact of the CO depletion on the line emission by truncating the $^{12}$CO
and $^{13}$CO abundance profiles at A$_\mathrm{v} = 5$ (setting the CO
abundance to zero for A$_\mathrm{v} \geq 5$; Model \#1b).  This is only a
rough approximation, however, it allows us to assess the impact of depletion
on the line emission.  Figure~\ref{fig:depletion} shows the $^{12}$CO and
$^{13}$CO $J = 3 \to 2 $ line profiles for Model \#1a and Model \#1b; the
resulting line-integrated intensities for the observed transitions are
summarized in Table~\ref{table:intensity_comparison}.  The line-integrated
intensities are smaller by 10$-$20\% in the model with depletion, mainly
because the line profiles are narrower.  We confirm that depletion plays a
negligible role for low$-J$ $^{12}$CO emission and note a larger impact on
both $^{13}$CO transitions (20\%) and the $^{12}$CO $J = 3 \to 2 $ transition
(5\%). However, the effect of depletion is much smaller than the factor of 2
to 3 discrepancy noted between the observations and the model line intensities
for $^{13}$CO. Thus, depletion of CO cannot solely account for the low
integrated intensity observed for the latter transitions.  In the following
models (Model \#2b, \#3a, and \#3b) we account for depletion in the same way
as done in Model \#1b.

\subsection{Polycyclic Aromatic Hydrocarbons}

Polycyclic Aromatic Hydrocarbons (PAHs) are known to play an important role in
the carbon chemistry, increasing the C column density and decreasing the CO
abundance in the cloud surface \citep{LeppDalgarno88,BakesTielens98}.
Following \citet{Kaufman99} we included PAHs with an initial abundance of
$1\times10^{-7}$ relative to H nuclei and assuming a MRN \citep{Mathis77} size
distribution for PAHs containing between 30 and 1500 carbon atoms.  We
consider neutral PAHs in our chemical network, their singly charged variants
(PAH$^+$, PAH$^-$), as well as PAHs with carbon and hydrogen ions adsorbed on
their surface (PAHC$^+$, PAHH$^+$).  We use a list of reactions provided by M.
Kaufman (private communication) and calculate the reaction rates from the work
of \citet{Draine87} and \citet{BakesTielens94}.  A comparison of the C$^+$,
C$^0$, and CO abundance profiles and the gas temperature in the model with and
without PAHs is shown in Figure~\ref{figure:2}.  We note that the results are
relatively insensitive to the initial PAH abundance as long as PAHs are
included.  Models with a factor of 4 larger PAH abundance give similar
line-integrated intensities (to within 10 \%).  One of the main impacts of
PAHs is that the C$^{+}$ layer at the cloud surface becomes thinner,
increasing the total atomic carbon column density.  Close to the surface of
the cloud PAH$^-$ neutralizes with C$^+$, while CO is formed somewhat deeper
into the cloud (A$_\mathrm{V} >$ 0.4).  The larger neutral carbon column
density (at A$_\mathrm{V} <$ 0.5; Figure~\ref{figure:2}) and the higher gas
temperature (A$_\mathrm{V} <$ 0.2) gives a stronger [C\,{\sc i}] emission
compared to the model without PAHs.  The CO line intensity is reduced in
Model \#2b because the emission traces a slightly colder region somewhat
deeper into the cloud.  The lower CO abundance at A$_\mathrm{V} <$ 0.4 in
Model \#2b also removes the self-absorption in the $^{12}$CO $J = 2\to1$ line
toward the cloud center and reduces it for the $^{12}$CO $J = 3\to2$ line.

Model \#2b gives [C\,{\sc i}]/CO line ratios consistent with the observations
and removes the $^{12}$CO $J=2 \to 1$ self-absorption noted in Model \#1b.
The $^{13}$CO line intensities are significantly reduced in the model with PAH
but are still a factor of $1.5-2.5$ larger than the observations. The [C\,{\sc
  i}] intensity is also overestimated by a factor of 1.5.

\subsection{Variations of the Radiation Field}
\label{sec:Radiation_Field}

\begin{figure*}[t]
  \centering \includegraphics[width=1.4\textwidth, angle=0]{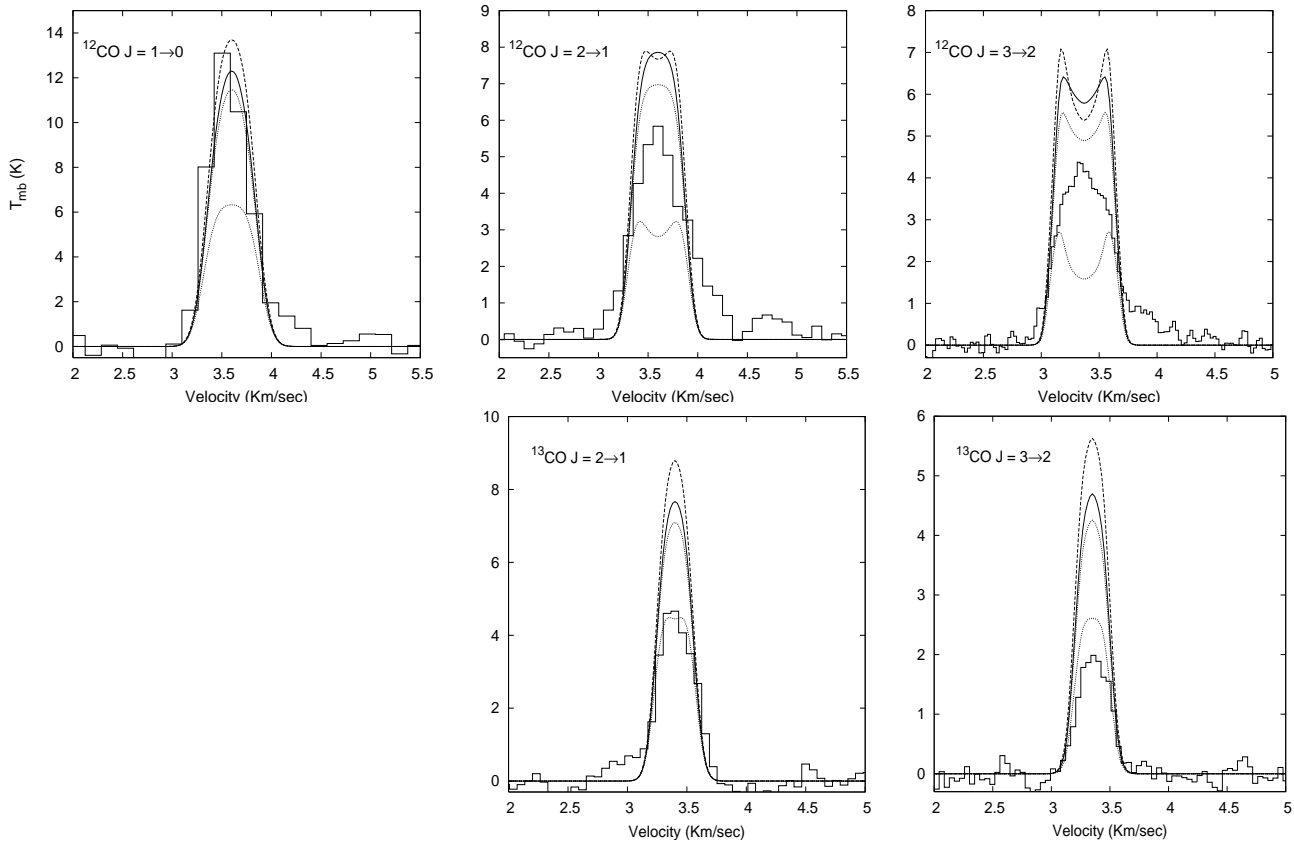}
      \caption{Observed  line profiles compared with  Model \#1b (dashed),
        \#2b (solid),\#3a (thick dotted), and \#3b (thin dotted).  The
        transitions are indicated in the upper left corner of each panel.}
        \label{figure:profile_triple}
\end{figure*}

\begin{figure*}[t]
         \centering
         \includegraphics[width=1.1\textwidth]{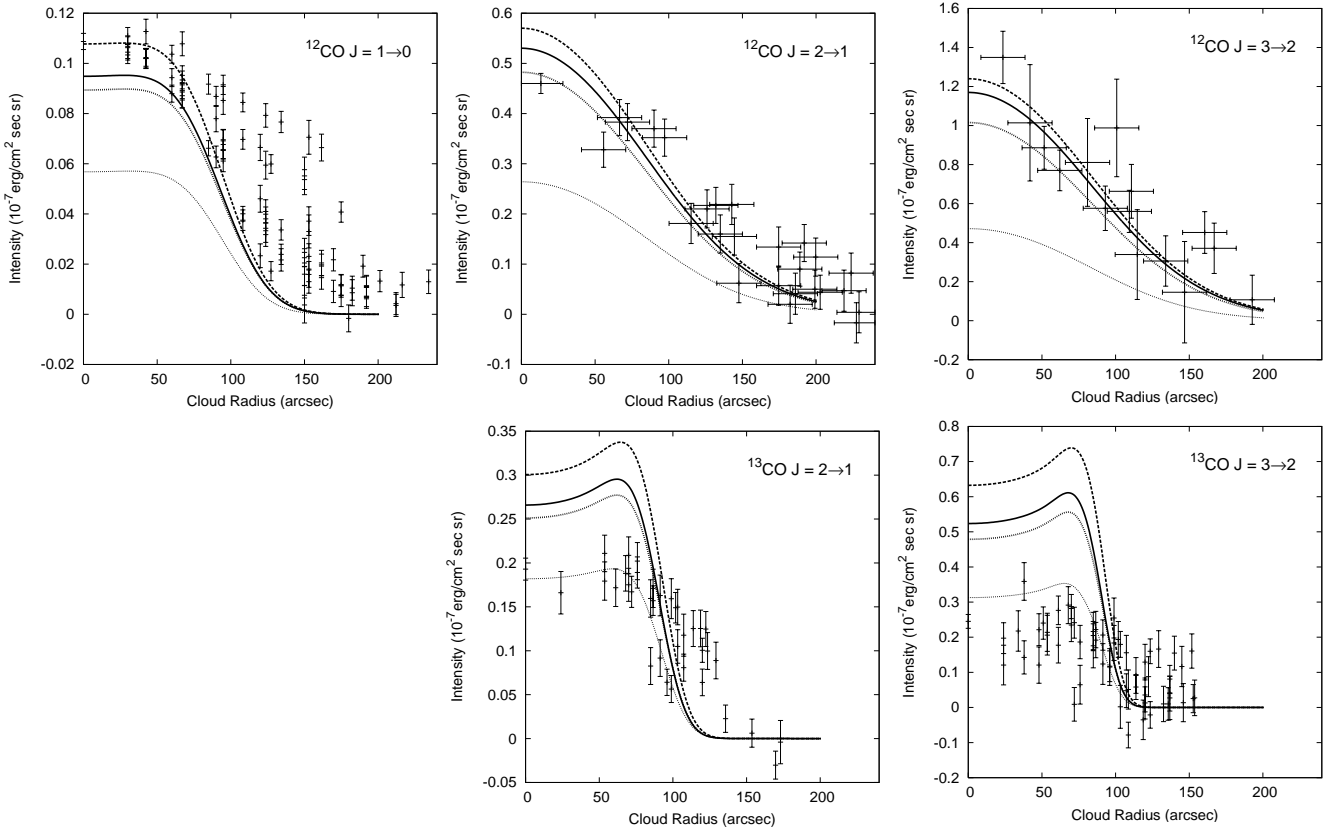}
      \caption{Line-integrated intensities as a
        function of cloud radius. The transitions are indicated in the upper
        right corner of each panel. The observations (error bars) are compared
        with predictions for Model \#1b (dashed), \#2b (solid),\#3a (thick
        dotted), and \#3b (thin dotted).  Horizontal error bars in the
        $^{12}$CO $J = 2\to1$ and $J = 3\to2$ emission correspond to the
        uncertainly in the correction for the receiver beam offset for the
        KOSMA observations (see Section \ref{sec:observations}).  We excluded
        the lower half of the Barnard~68 data from comparison to the CO data
        because of the significant deviation from the spherical symmetry,
        visible in the dust extinction maps \citep{Alves01}.  }
\label{figure:cloudprofile}
   \end{figure*}

\begin{table*}[t]
\caption{Predicted line-integrated intensities toward Barnard 68 center}             
\label{tab:predictions}
\centering                  
\begin{tabular}{l l c c c c}  
\hline\hline                 
Species Transition & Resolution -  Instrument  & \multicolumn{4}{c}{Intensity (10$^{-7}$ erg~s$^{-1}$~cm$^{-2}$~sr$^{-1}$)}    \\

&& Model \#1a  &Model \#1b  & Model \#2b & Model \#3a\\

\hline                        
\\[-0.25cm]
[C\,{\sc ii}] $^2\mathrm{P}_{3/2}$$\to$$^2\mathrm{P}_{1/2}$ &
18\arcsec-GREAT-SOFIA & 9.49 & 9.49 & 4.77 & 2.44\\ 

& 13\arcsec-HIFI-Herschel & 9.47  & 9.47  & 4.76& 2.43\\

[C\,{\sc i}] $^3\mathrm{P}_2\to^3\mathrm{P}_1$  & 25\arcsec-NANTEN2 &
2.83 & 2.86 & 6.65 & 5.43 \\
& 25\arcsec-HIFI-Herschel &  &  &  & \\

& 8\arcsec-APEX & 2.83 &  2.86 & 6.62 & 5.40 \\

$^{12}$CO $J = 5 \to 4$  & 39\arcsec-HIFI-Herschel & 2.49 & 2.38 & 1.60 & 1.37\\ 
& 56\arcsec-CASIMIR-SOFIA & 2.51 & 2.39 & 1.61 & 1.37\\

$^{12}$CO $J = 4 \to 3$ & 13\arcsec-APEX  & 2.22 & 2.12  & 1.83 & 1.56    \\
 & 45\arcsec-NANTEN2  & 2.23  & 2.13 & 1.84 & 1.56  \\

\hline                                   
\end{tabular}
\end{table*}

\begin{figure*}[t]
  \centering \includegraphics[width=1.4\textwidth]{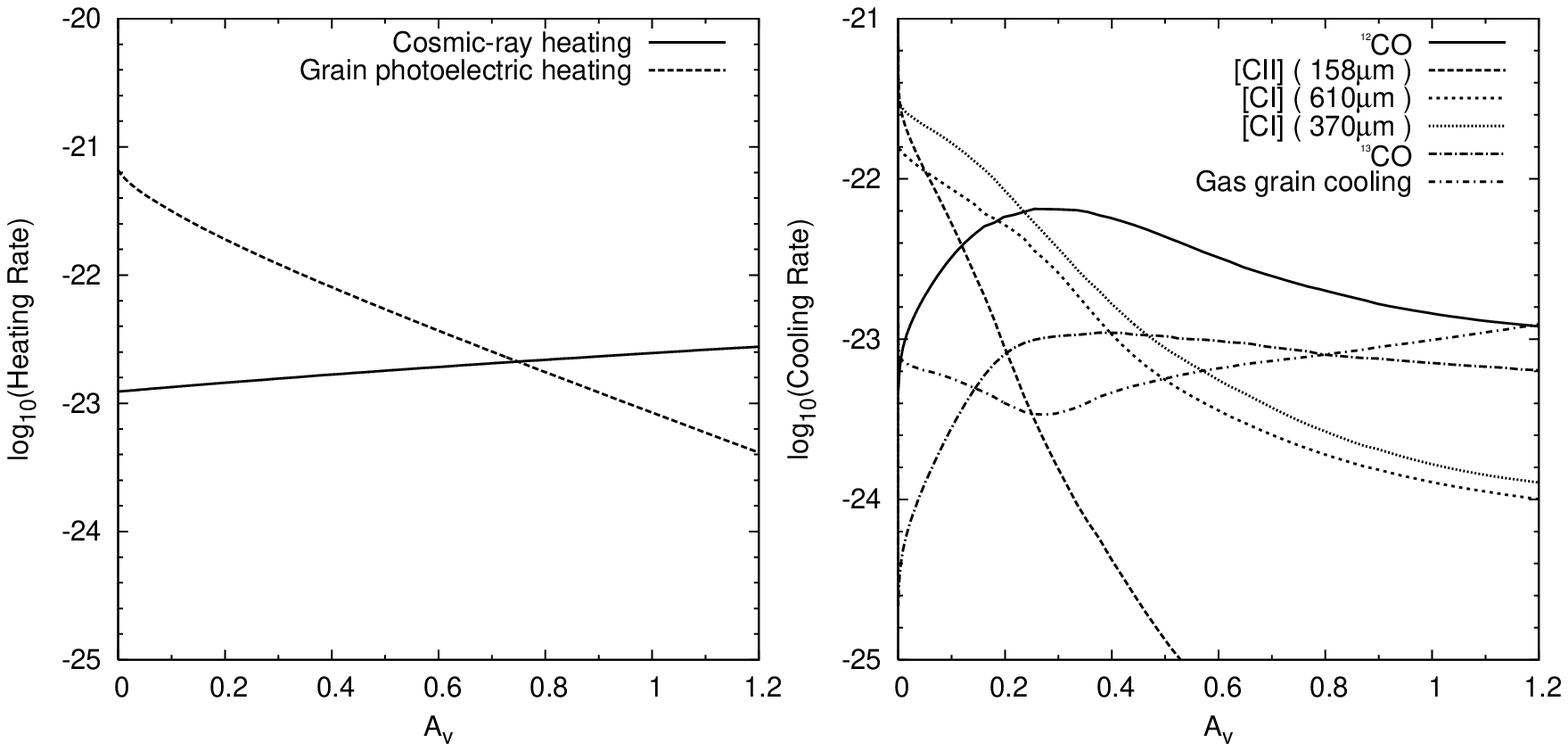}
      \caption{Dominant heating and cooling processes for Model \#3a ($\chi =
        0.75\chi_0$). The heating and cooling rates are in units of erg
        s$^{-1}$ cm$^{-3}$.}
     \label{figure:cooling}
\end{figure*}

Generally, the line emission in the PDR model is reduced if a weaker external
FUV radiation field is assumed. We explore the impact of lower external
radiation field intensity as a next step. According to \citet{Bergin06}, the
low $^{13}$CO line intensities suggest that Barnard 68 is exposed to a FUV
radiation field of rather low intensity, corresponding to $\chi = 0.12\chi_0$
(0.2 in units of the \citet{Habing68} field).  We run two additional models,
the first with a somewhat smaller $\chi$ = 0.75$\chi_0$ (Model \#3a) and a
second with an intensity following \citet{Bergin06} ($\chi$ = 0.12$\chi_0$;
Model \#3b).  All other model parameters are the same as for Model \#2b. The
observed CO line profiles and the model line profiles are shown in
Figure~\ref{figure:profile_triple}, and the radial profiles of the
line-integrated intensity are given in Figure~\ref{figure:cloudprofile}.  
  A comparison between the observed intensity profiles and the model is shown
  in Table~\ref{tab:chi2_v2}. In order to quantify their difference, we
  calculate the average ratio of the model simulation to the observations for
  each observed position within $r < 100$\arcsec.

\begin{table*}[t]
\caption{Model comparison with observed cloud intensity profiles}  
\label{tab:chi2_v2}
\centering          
\begin{tabular}{l c c  c c  }
\hline\hline

Species Transition &  \multicolumn{4}{c}{$N^{-1}\Sigma_i^{N} \,I_\mathrm{model}/I_\mathrm{obs}^{a}$}  \\    
 &  Model \#1b &   Model \#2b  & Model  \#3a  & Model \#3b \\    
\hline    
$^{12}$CO $J = 1 \to 0$& 0.95 & 0.82 & 0.78 & 0.50 \\
$^{12}$CO $J = 2 \to 1$& 1.10 & 0.98 & 0.90 & 0.48 \\
$^{12}$CO $J = 3 \to 2$& 1.10 & 1.00 & 0.87 & 0.40 \\
$^{13}$CO $J = 2 \to 1$& 1.78 & 1.49 & 1.40 & 0.98 \\
$^{13}$CO $J = 3 \to 2$& 3.28 & 2.60 & 2.39 & 1.50 \\

\hline                 
\multicolumn{5}{l}{$^{a}$ Sum over all positions with  $r<100\arcsec$.}              
\\[-0.25cm]
\end{tabular}
\end{table*}

While the line intensities of both species, [C\,{\sc i}] and CO, are reduced
for a smaller $\chi$, the [C\,{\sc i}] emission is more sensitive to the
intensity of the external FUV field than the CO rotational transitions.  A
weaker FUV radiation field significantly decreases the gas temperature at the
cloud surface (Figure~\ref{figure:2}a) but does not change the C$^{0}$ density
appreciably (Figure~\ref{figure:2}b). Generally, the gas temperature in the
models is below 23.6~K, the excitation energy of the carbon $^3$P$_1$ level
above the ground state.  In this domain the [C\,{\sc i}] line intensity is
very sensitive to the temperature and thus to $\chi$. In contrast, reducing
the FUV intensity results only in small changes of the gas temperature in the
region where CO is the most abundant carbon species and thus has a
correspondingly small impact on the optically thick CO lines. The exception is
Model \# 3b with an extremely low FUV radiation field of $\chi$ =
0.12$\chi_0$. In this case the cosmic-ray heating prevails already at
A$_\mathrm{v} \gtrsim 0.4$, with even {\it lower} temperatures closer to the
surface (Figure~\ref{figure:2}). 

The strength of the FUV radiation field also influences the self-absorption in
the model $^{12}$CO $J = 2 \to 1$ and $J = 3 \to 2$ line profiles because the
latter depends on the gas temperature at the cloud surface.  As these
transitions are sub-thermally excited at the lower-density gas near the cloud
edge, the reduction of the gas temperature in the low-FUV model results in
stronger self-absorption (Figure~\ref{figure:profile_triple}; thin dotted
lines).

The lower FUV intensity significantly reduces the [C\,{\sc i}] and $^{13}$CO
emission. The model with $\chi$ = 0.12$\chi_0$ (Model \#3b) reproduces the
observed $^{13}$CO intensities and line-integrated intensity profiles to
within $\sim$40\%, confirming the results by \citet{Bergin06}. However, this
model cannot account for the observed [C\,{\sc i}] and $^{12}$CO emission,
which is underestimated by a factor of $\gtrsim$2.  Additionally, we note
that models with a low FUV radiation field produce clear self-absorption in
the $^{12}$CO $J = 2 \to 1$ and $J = 3 \to 2$ line profiles which is not
observed.  The model with a slightly reduced FUV radiation field ($\chi$ =
0.75$\chi_0$; Model \#3a) reduces the [C\,{\sc i}] and $^{13}$CO emission by
10\% compared to the $\chi$ = 1.0$\chi_0$ model.  While the $^{12}$CO model
intensity still marginally agrees with the data (within $\lesssim 30\%$), the
$^{13}$CO emission is significantly larger than observed. Any further
reduction brings the $^{13}$CO model intensity closer to the observations, but
at the same time gives [C\,{\sc i}] and $^{12}$CO intensities that are too
small.

We notice that the model predictions for the $^{12}$CO $J = 1 \to 0$
integrated intensity profile show larger discrepancies at radii between
90\arcsec~and 150\arcsec. This is possibly due to deviations from spherical
symmetry at angular scales $\lesssim 60$\arcsec~(note, for example, the
deviations of the iso-intensity contours from circular symmetry in
Figure~\ref{ci_swas}; recall that we excluded the southern half of Barnard~68
for the same reason, see Fig.\,~\ref{figure:cloudprofile}). These deviations
are therefore more noticeable in the high-angular resolution $^{12}$CO $J = 1
\to 0$ map than in the KOSMA maps (angular resolution 130\arcsec).  We checked
for any possible contribution from stray-emission from the Mopra error beam
($\sim$60\arcsec, \citealt{Ladd05}) and concluded that this cannot account for
a significant fraction of the excess emission observed at 90\arcsec~$\lesssim
r \lesssim$ 150\arcsec.  Also note that emission from a low-A$_\mathrm{v}$
region at $100 \lesssim r \lesssim 120$ (Figure~\ref{colden}) cannot account
for this excess (see discussion in Section~\ref{sec:reference-model}).

\section{Discussion}
\label{sec:Discussion}

\begin{table}
\caption{Line-integrated intensity averaged over the projected clump area for
  Model \#3a ($\chi=0.75\chi_0$).}    
\label{tab:line_cooling}
\centering                  
\begin{tabular}{l c}        
\hline\hline                 
Species Transition & Intensity  \\    
& \small (10$^{-7}$ erg~s$^{-1}$~cm$^{-2}$~sr$^{-1}$)     \\
\hline
[C\,{\sc ii}] $^2\mathrm{P}_{3/2}$$\to$$^2\mathrm{P}_{1/2}$ & 3.94 \\ 

[C\,{\sc i}] $^3\mathrm{P}_1\to^3\mathrm{P}_0$ & 3.76 \\

[C\,{\sc i}] $^3\mathrm{P}_2\to^3\mathrm{P}_1$ & 6.95 \\
$^{12}$CO $J = 3 \to 2$ & 1.19 \\
$^{12}$CO $J = 4 \to 3$ & 1.38  \\
$^{12}$CO $J = 5 \to 4$ & 1.13  \\
$^{13}$CO $J = 2 \to 1$ & 0.23\\
$^{13}$CO $J = 3 \to 2$ & 0.45 \\
$^{13}$CO $J = 4 \to 3$ & 0.30 \\
\hline    
\end{tabular}
\end{table}

\subsection{Influence of the radiation field}

We find that for the given density profile of Barnard~68 it is not possible to
find a PDR model that matches all observations.  The main difficulty is to
simultaneously match the $^{12}$CO and $^{13}$CO transitions.  The weak
$^{13}$CO lines suggest a low FUV intensity of $\chi = 0.12 \chi_0$, but
underestimate the [C\,{\sc i}] and $^{12}$CO intensity by up to a factor of
$\sim$2 and produce deep self-absorbed $^{12}$CO line profiles towards the
cloud center.  The $^{12}$CO and [C\,{\sc i}] intensities suggest a moderately
reduced FUV field $\chi = 1-0.75 \chi_0$.  However, this overestimates the
$^{13}$CO intensity by a factor of $1.5-2.5$.  A strength of the FUV field of
the order of the mean interstellar radiation field have been independently
inferred from observations at 7$\mu m$ and 90$\mu m$ by \citealt[][]{Galli02}
($\sim$2.5\,G$_{0}$, corresponding to $\sim1.5 \chi_0$), arguing against a
substantial reduction of the FUV field at the location of Barnard\,68.  A
smaller impact is noticed by including PAHs in the chemical network of the
model.  The [C\,{\sc i}] intensity is increased by 60\%, while the CO
intensity is reduced by 20\%.  PAHs help to remove or reduce the
self-absorption and make the line profiles more consistent with the
observations. Finally, we note that depletion affects the observed transitions
only to a small degree.

Observations of higher$-J$ CO, [C\,{\sc i}] $^3$P$_2$$\to$$^3$P$_1$, and
[C\,{\sc ii}] transitions are useful to discriminate between the impact of
relevant model parameters and whether the low $^{13}$CO intensities indeed
result from a low FUV field. They will also allow to better assess the impact
of PAHs and/or depletion due to freezing on dust even at the cloud surface.
Predictions for Model \#1b, \#2b, \#3a, and \#3b are listed in
Table~\ref{tab:predictions} for observations with present and future
observatories.  The impact of PAHs is a reduction of the [C\,{\sc ii}]
intensity by a factor of 2, while the [C\,{\sc i}] $^3$P$_2$$\to$$^3$P$_1$
intensity is increased by the same factor.  Because of their high excitation
energy, both the [C\,{\sc ii}] and [C\,{\sc i}] $^3$P$_2$$\to$$^3$P$_1$
transitions are {\it very} sensitive to the FUV field.  They are reduced by up
to a factor of 0.5 when the FUV field is reduced to $\chi = 0.75 \chi_0$, and
by as much as a factor of 1.4$\times$10$^{-4}$ for $\chi = 0.12 \chi_0$. The
$^{12}$CO $J = 5 \to 4$ and $J =4 \to 3$ line intensities are more moderately
reduced.  Together, the inclusion of PAHs and a somewhat lower FUV field
($\chi = 0.75 \chi_0$) reduce them by 40\%.

The detection of the $^{12}$CO $J = 4\to 3$ (APEX, NANTEN2) and $^{12}$CO $J =
5\to 4$ (HIFI Herschel, GREAT SOFIA) transitions can be obtained within a few
minutes or less of observing time. A substantially longer integration time is
required for the [C\,{\sc ii}] $^2\mathrm{P}_{3/2}$$\to$$^2\mathrm{P}_{1/2}$
transitions. The detection of the [C\,{\sc i}] $^3$P$_2$$\to$$^3$P$_1$
transition might also prove very difficult from ground based observatories.
However, with HIFI-Herschel, the [C\,{\sc i}] $^3$P$_2$$\to$$^3$P$_1$
transition can be detected in less than a minute.

\subsection{Heating and cooling}

In Figure \ref{figure:cooling} we show the heating and cooling for different
processes as a function of the visual extinction A$_\mathrm{v}$ (Model \#3a,
$\chi$ = 0.75$\chi_0$).  Photoelectric heating dominates the gas heating at
A$_\mathrm{v} \lesssim$ 0.8, being replaced by cosmic-ray heating at larger
depths. The gas cooling is governed by the [C\,{\sc i}]
$^3\mathrm{P}_{1}$$\to$$^3\mathrm{P}_{0}$ and
$^3\mathrm{P}_{2}$$\to$$^3\mathrm{P}_{1}$ fine structure transitions for
A$_\mathrm{v} < $ 0.2, and by the $^{12}$CO (and $^{13}$CO) rotational
transitions for 0.2 $<$ A$_\mathrm{v} <$ 1.2.  Gas-grain coupling dominates at
A$_\mathrm{v} >$ 1.2. It is interesting to note that the [C\,{\sc ii}]
transition does not dominate the cooling in this case, even very close to the
cloud surface (C$^{+}$ dominates only at A$_\mathrm{v} \lesssim$ 0.05).

Table~\ref{tab:line_cooling} lists  the line emission averaged over the
  projected cross section of the clump for Model \#3a ($\chi=0.75\chi_0$).
The [C\,{\sc i}] fine structure transitions dominate with a total intensity of
1.1$\times$10$^{-6}$ erg~s$^{-1}$~cm$^{-2}$~sr$^{-1}$, followed by the
$^{12}$CO and $^{13}$CO rotational transitions with a total intensity of
4.7$\times$10$^{-7}$ erg~s$^{-1}$~cm$^{-2}$~sr$^{-1}$. The [C\,{\sc ii}]
$^2$P$_{3/2}$$\to$$^2\mathrm{P}_{1/2}$ fine structure transition contributes
with a total intensity of 3.9$\times$ 10$^{-7}$ erg~s$^{-1}$
cm$^{-2}$~sr$^{-1}$. This relatively low value is attributed to the low FUV
radiation field of Model \#3a and to the presence of PAHs in the chemical
network of the model, which significantly decreases the abundance of C$^+$
close to the surface of the cloud.

\section{Summary  and Conclusions}
\label{sec:conclusions}

We present simulations with a spherical PDR code to model the line emission in
Barnard~68. Its approximately spherical geometry makes Barnard~68 an ideal
test for such a spherically-symmetric PDR model.  We compare the model results
with the line profiles and the azimuthally averaged intensity profiles of the
three lowest rotational transitions of $^{12}$CO, $^{13}$CO $J = 3 \to 2 $, $J
= 2 \to 1 $, and the spatially and spectrally unresolved [C\,{\sc i}]
$^3\mathrm{P}_{1}$$\to$$^3\mathrm{P}_{0}$ transition.  This is the first time
that a spherical PDR code was used to model the line emission in a low-FUV
cloud in such a detail.

We have tested the impact of PAHs in the chemical network of the model, the
FUV radiation field intensity, and considered the effects of depletion.  The
impact on the emerging $^{12}$CO, $^{13}$CO, [C\,{\sc i}], and [C\,{\sc ii}]
line emission is quantified.  We find evidence that PAHs play a role in the
chemical network.  This results in a larger C$^{0}$ layer, while the C$^{+}$
layer is very thin, making the [C\,{\sc ii}] emission very faint and difficult
to discriminate from emission from the ambient Warm Ionized Medium (WIM) in
spectrally unresolved observations.  Depletion is not likely to play an
important role in the observed transitions.

We find it difficult to simultaneously model the observed $^{12}$CO and
$^{13}$CO emission, with residual differences up to a factor of 2 between PDR
models and observed line intensities.  The weak $^{13}$CO line intensity seems
to require a steep drop in either the excitation conditions or in the
abundances in a layer between the regions which dominate the $^{12}$CO and
$^{13}$CO emission, respectively.  A model with a low external FUV field
\citep[see also ][]{Bergin06} is clearly incompatible with the $^{12}$CO and
[C\,{\sc i}] observations.   In order to reduce the $^{13}$CO line
  emission, a shift of the $^{13}$CO abundance to a region where cosmic rays
  dominate the gas heating (A$_\mathrm{v}>0.8$) is required.  However,
  isotope-selective reactions produce the opposite effect and a $^{13}$CO
  abundance corresponding to the $^{13}$C isotope abundance is reached in a
  region which is even closer to the cloud surface than for $^{12}$CO
  (Fig.~\ref{figure:2}).  Consequently, the low$-J$ $^{13}$CO emission is
  optically thick at A$_\mathrm{v}=0.3-0.5$ where the gas heating is still
  dominated by the external FUV radiation field (Fig.~\ref{figure:cooling}).
  Including PAHs in the model reduces the gas temperature in the region where
  $^{13}$CO is emitted, but this effect turns out to be insufficient.  The
  $^{13}$CO intensity could be significantly reduced if the gas-grain
  collisions dominate the cooling closer to the surface and the dust
  temperature was close to $\sim$10\,K.  However, a confirmation of the low
  observed $^{13}$CO intensity is also highly desirable.  Together with
  observations of mid$-J$ $^{12}$CO, [C\,{\sc i}], and [C\,{\sc ii}]
  transitions with present and future telescopes (NANTEN2, APEX, SOFIA,
  Herschel), this will allow a test of our results and assess the importance
  of different model parameters.

\begin{acknowledgements}
  We want to thank Edwin Bergin for providing us with the CSO $^{13}$CO
  observations, and Michael Kaufman for his help to include PAHs in the
  chemical network of our PDR model.  This work is supported by the
  \emph{Deut\-sche For\-schungs\-ge\-mein\-schaft, DFG\/} via Grant SFB 494.
  J.L.P. was supported for this research through a stipend from the
  International Max Planck Research School (IMPRS) for Radio and Infrared
  Astronomy at the University of Bonn and Cologne. The KOSMA telescope on
  Gonergrat is joinly operated by the Universities of Cologne and Bonn, and
  supported by a special fund from the Land Nordrhein-Westfalen. The
  observatory is administrated by the Internationale Stiftung Hochalpine
  Forschungstationen Jungfraujoch und Gornergrat. The Mopra radio telescope is
  part of the Australia Telescope which is funded by the Commonwealth of
  Australia for operation as a National Facility managed by CSIRO.
\end{acknowledgements}

\bibliographystyle{aa}
\bibliography{../latex/papers}

\end{document}